\documentclass[aps,preprint]{revtex4}%
\usepackage{amsfonts}
\usepackage{amsmath}
\usepackage{amssymb}
\usepackage{graphicx}%
\providecommand{\U}[1]{\protect\rule{.1in}{.1in}}

\begin{document}
\author{S A Bruce and P C Minning}
\affiliation{Department of Physics, University of Concepcion, P.O. Box 160C, Concepcion, Chile}
\title{Supersymmetry, quark confinement and the harmonic oscillator}
\date{22/10/07}

\begin{abstract}
We study some quantum systems described by noncanonical commutation relations
formally expressed as $\left[  \widehat{q},\widehat{p}\right]  =i\hbar\left(
\widehat{I}+\widehat{\chi}\widehat{H}_{\text{HO}}\right)  $, where
$\widehat{H}_{\text{HO}}$ is the associated (harmonic oscillator-like)
Hamiltonian of the system, and $\widehat{\chi}$ is a Hermitian (constant)
operator, i.e. $\left[  \widehat{H}_{\text{HO}},\widehat{\chi}\right]
=\widehat{0}$ . In passing, we also consider a simple ($\widehat{\chi
}=\widehat{0}$ canonical) model, in the framework of a relativistic
Klein-Gordon-like wave equation.

PACS\ 03.65.Pm - Relativistic wave equations,

PACS 02.20.Sv \ - Lie algebras of Lie groups,

PACS 12.60.Jv \ - \ Supersymmetric models.

\end{abstract}
\maketitle

\section{Introduction}

This article deals with some possible applications of certain harmonic
oscillator-like models to particle physics. From the beginning we should
mention that the spirit of this paper is rather formal and analytical. Since
the beginning of modern physics, an enormous amount of work has been produced
regarding the harmonic oscillator: its various forms and applications to
quantum and classical mechanics. It is clear to everyone that the concept of
harmonic oscillator is a fundamental source to understand many concrete
problems in physics \cite{MH1}. Therefore we content ourselves by presenting
certain mathematical relations between deep-seated symmetries in which some
specific kinds of harmonic oscillators are involved. Hence, it is not
presently our goal to give explicit numerical solutions to the problems we
address, since they are available (under various concepts) in the literature.

The main premise of this article is to study some possible quantum systems
with noncanonical commutation relations of the form%
\begin{equation}
\left[  \widehat{q}^{i},\widehat{p}^{j}\right]  =i\hbar\delta^{ij}\left(
\widehat{I}+\widehat{\chi}\widehat{H}_{\text{HO}}\right)  \ , \label{i1}%
\end{equation}
where $\widehat{H}_{\text{HO}}$ is the harmonic oscillator-like Hamiltonian
(whose specific form depends on the concrete nature of the treated system) and
$\widehat{\chi}$ is a Hermitian (constant) operator, i.e. $\left[  \widehat
{H}_{\text{HO}},\widehat{\chi}\right]  =\widehat{0}$.

In section II we present a particular type of harmonic oscillator in $\left(
1+1\right)  $-dimensions as an underlying device in its connection with
supersymmetry and quark confinement. Sections III concerns some corresponding
generalizations to $\left(  3+1\right)  $-dimensions.

\section{A harmonic oscillator in $\left(  1+1\right)  $-dimensions}

We begin by considering a quantum system consisting of a particle of mass
$m_{0}$ moving in a $\left(  1+1\right)  $-dimensional space-time described by
the (noncanonical) \textquotedblleft coordinate and momentum\textquotedblright%
\ realizations%
\begin{equation}
\widehat{Q}\equiv\frac{c}{\omega}\sqrt{\frac{1}{2\hbar m_{0}\omega}}\left(
m_{0}\omega\sigma_{2}\widehat{q}+\sigma_{1}\widehat{p}\right)
\ ,\ \ \ \ \ \ \ \ \widehat{P}\equiv\frac{\hbar\omega}{c}\sqrt{\frac{1}{2\hbar
m_{0}\omega}}\left(  \sigma_{2}\widehat{p}-m_{0}\omega\sigma_{1}\widehat
{q}\right)  \ , \label{i10}%
\end{equation}
where $\sigma_{j}$ are Pauli matrices. In Eq.(\ref{i10}) the operators
$\widehat{q}$, $\widehat{p}$ are the usual canonical coordinates, so that
$\left[  \widehat{q},\widehat{p}\right]  =i\hbar\widehat{I}$. Here
$\widehat{Q}$ and $\widehat{P}$ are Hermitian and formally traceless
operators: $\widehat{Q}^{\dag}=\widehat{Q}$, $\ \widehat{P}^{\dag}=\widehat
{P}$, and Tr$\left(  \widehat{Q}\right)  =$Tr$\left(  \widehat{P}\right)  =0$.
Next, we determine $\widehat{Q}\widehat{P}$ and $\widehat{P}\widehat{Q}$:
\begin{align}
\widehat{Q}\widehat{P}  &  =\frac{1}{2}\left(  \widehat{q}\widehat{p}%
-\widehat{p}\widehat{q}\right)  +\frac{1}{\omega}\sigma_{1}\sigma_{2}\left(
\frac{\widehat{p}^{2}}{2m_{0}}+\frac{m_{0}\omega^{2}}{2}\widehat{q}%
^{2}\right)  \equiv\frac{i\hbar}{2}\widehat{I}+\frac{i}{\omega}\sigma
_{3}\widehat{H}_{\text{HO}}=\frac{i\sigma_{3}}{\omega}\widehat{H}_{\text{SS}%
}\ ,\\
\widehat{P}\widehat{Q}  &  =-\left(  \frac{i\hbar}{2}\widehat{I}+\frac
{i}{\omega}\sigma_{3}\widehat{H}_{\text{HO}}\right)  =-i\frac{\sigma_{3}%
}{\omega}\widehat{H}_{\text{SS}}\ ,
\end{align}
in which $\widehat{H}_{\text{HO}}\equiv\left(  1/2\right)  \left(  \widehat
{p}^{2}/m_{0}+m_{0}\omega^{2}\widehat{q}^{2}\right)  $, and $\widehat
{H}_{\text{SS}}$ is the well known supersymmetric Hamiltonian for the harmonic
oscillator \cite{LA}. Thus%
\begin{equation}
\left[  \widehat{Q},\widehat{P}\right]  =i\hbar\left(  \widehat{I}+\left(
\frac{2\sigma_{3}}{\hbar\omega}\right)  \widehat{H}_{\text{HO}}\right)
=-\frac{2i\sigma_{3}}{\omega}\widehat{H}_{\text{SS}}%
\ ,\ \ \ \ \ \ \ \ \ \left\{  \widehat{Q},\widehat{P}\right\}  =\widehat{0}\ ,
\label{i20}%
\end{equation}
where $\widehat{\chi}=2\sigma_{3}/\hbar\omega$ , and $\sigma_{3}$ may be
interpreted as a `charge operator'. Furthermore%
\begin{align}
\widehat{Q}^{2}  &  =\frac{c}{\hbar\omega^{3}}\left(  \frac{\widehat{p}^{2}%
}{2m_{0}}+\frac{m_{0}\omega^{2}}{2}\widehat{q}^{2}+\frac{\hbar\omega}{2}%
\sigma_{3}\right)  =\frac{c}{\hbar}\left(  \frac{1}{\omega}\right)
^{2}\left(  \frac{1}{\omega}\right)  \widehat{H}_{\text{SS}}\ ,\label{i15}\\
\widehat{P}^{2}  &  =\frac{\hbar\omega}{c^{2}}\left(  \frac{\widehat{p}^{2}%
}{2m_{0}}+\frac{m_{0}\omega^{2}}{2}\widehat{q}^{2}+\frac{\hbar\omega}{2}%
\sigma_{3}\right)  =\frac{\hbar\omega}{c^{2}}\widehat{H}_{\text{SS}%
}\ ,\nonumber
\end{align}
from which we get
\begin{equation}
\widehat{H}_{\text{SS}}\equiv\frac{1}{2}\hbar\omega\left\{  \left(  \frac
{c}{\hbar\omega}\right)  ^{2}\widehat{P}^{2}+\left(  \frac{\omega}{c}\right)
^{2}\widehat{Q}^{2}\right\}  =\frac{\widehat{p}^{2}}{2m_{0}}+\frac{1}{2}%
m_{0}\omega^{2}\widehat{q}^{2}+\frac{\hbar\omega}{2}\sigma_{3}\ .
\end{equation}
Additionally, from Eqs.(\ref{i15}) it follows that%
\begin{equation}
\left[  \widehat{H}_{\text{SS}},\widehat{Q}\right]  =\left[  \widehat{Q}%
^{2},\widehat{Q}\right]  =\widehat{0}\ ,\ \ \ \ \ \ \ \ \ \ \left[
\widehat{H}_{\text{SS}},\widehat{P}\right]  =\left[  \widehat{P}^{2}%
,\widehat{P}\right]  =\widehat{0}\ , \label{i30}%
\end{equation}
which closes the algebra and confirms that the (positive-definite) Hamiltonian
$H_{\text{SS}}$ is invariant under this supersymmetry.

Finally%
\begin{equation}
\left[  \widehat{H}_{\text{HO}},\widehat{Q}\right]  =-i\frac{c^{2}}{\omega
}\widehat{P}\ ,\ \ \ \ \ \ \ \ \ \ \left[  \widehat{H}_{\text{HO}},\widehat
{P}\right]  =i\omega\left(  \frac{\hbar\omega}{c}\right)  ^{2}\widehat{Q}.
\end{equation}

From Eq.(\ref{i20}), the operator $\widehat{H}$ $\equiv\left(  2\sigma
_{3}/\hbar\omega\right)  \widehat{H}_{\text{HO}}$ incorporates negative-energy
eigenvalues into the system. In fact, the structure of this equation reminds
us of the Zitterbewegung\textit{\ }(ZB)\textit{\ }phenomenon (as a result of
the eventual interference between positive and negative-energy eigenstates)
extensively studied by Barut \cite{BA} amongst others. On the other hand, note
that, from the point of view of SUSYQM, $\widehat{Q}$ and $\widehat{P}$ are
proportional to SUSY charges \cite{WI}.

Thus we have encountered three non-equivalent Hamiltonians: $\widehat
{H}_{\text{HO}}$,$\ \widehat{H}_{\text{SS}}$ (boson-fermion symmetry),$\ $and
$\widehat{H}$ (particle-antiparticle symmetry). In principle each one of them
refers to a different (in nature) quantum system.

\subsection{Modeling `quark confinement' in $\left(  1+1\right)  $-dimensions}

In $\left(  1+1\right)  $-dimensions, some authors
\cite{SY,SA1,SA2,MO,KKE,AKE,YO} have argued that a modification of the usual
canonical coordinates $\widehat{q}$, $\widehat{p}$ into a set of noncanonical
coordinates (say) $\widehat{Q}$,$\ \widehat{P}$, could lead to a modification
of the usual commutation relation, similar to the first Eq.(\ref{i20}), into
the form%
\begin{equation}
\left[  \widehat{Q},\widehat{P}\right]  =i\hbar\left(  \widehat{I}%
+\chi\widehat{H}\right)  \ , \label{q10}%
\end{equation}
where $\chi$ is just a parameter with dimensions of $\left[  \text{Energy}%
\right]  ^{-1}$. With this (oversimplified) scheme it is possible to attempt
to give some hints in order to describe new phenomena, which appear, for
instance, in high energy physics and mesoscopic systems. In Ref.\cite{SY} it
is shown that for a quantum system fulfilling commutation relation
(\ref{q10}), space discreteness is compatible with Lorentz transformations.
This fact was explicitly related to atomic phenomena. In Refs.\cite{SA1,SA2} a
mass-spectrum for a subset of elementary particles was obtained from
Eq.(\ref{q10}), with $\widehat{H}$ the Hamiltonian of the harmonic oscillator,
and applied for energies of the order of GeV-TeV ($10^{10}\ $-$\ 10^{12}$ eV).
In Ref.\cite{MO} it was also shown that, for the free-particle Hamiltonian,
Eq.(\ref{q10}) reproduces space quantization. This result can be related to
quark confinement phenomena. Qualitatively, in Eq.(\ref{q10}) as
$\chi\rightarrow\infty$, the theory becomes `asymptotically free' since
$\widehat{H}\varpropto\widehat{\mathbf{p}}^{2}$ . On the other hand, if
$\chi\rightarrow0$, $\widehat{H}\varpropto\widehat{\mathbf{q}}^{2}$ , the
parameter $1/\chi$ corresponds to a strong coupling constant: `quark
confinement in $\left(  1+1\right)  $-dimensions'. Furthermore, in
Refs.\cite{KKE,AKE} mathematical aspects of Eq.(\ref{q10}) were studied. In
Ref.\cite{YO} it was found that charge discreteness in mesoscopic circuits can
be mathematically formulated with commutation relations similar to that of
Eq.(\ref{q10}) between charge and current. This theory becomes related to the
descriptions of phenomena like persistent current in a ring of inductance $L$,
i.e., Coulomb blockage phenomena in a pure capacitor-design.

\section{Some generalizations in $\left(  3+1\right)  $-dimensions}

At least three straightforward (non-equivalent) quantum models can be
visualized in $\left(  3+1\right)  $-dimensions.

\subsection{Supersymmetry algebra}

By somehow mimicking the $\left(  1+1\right)  $-dimensional case, we address a
quantum system consisting of a spin-$1/2$ particle of mass $m_{0}$ moving in a
$\left(  3+1\right)  $-dimensional space-time described by the noncanonical
\textquotedblleft coordinate and momentum\textquotedblright\ Hermitian
operators%
\begin{equation}
\widehat{Q}\equiv\frac{c}{\omega}\sqrt{\frac{1}{6\hbar m_{0}\omega}}\left(
m_{0}\omega\mathbf{\alpha\cdot}\widehat{\mathbf{q}}+i\beta\mathbf{\alpha\cdot
}\widehat{\mathbf{p}}\right)  \ ,\ \ \ \ \ \widehat{P}\equiv\frac{\hbar\omega
}{c}\sqrt{\frac{1}{6\hbar m_{0}\omega}}\left(  \mathbf{\alpha\cdot}%
\widehat{\mathbf{p}}-m_{0}\omega i\beta\mathbf{\alpha\cdot}\widehat
{\mathbf{q}}\right)  \ , \label{g10}%
\end{equation}
where $\beta,\alpha_{j}$ $\left(  j=1,2,3\right)  $ are Dirac matrices. The
operators $\widehat{q}^{i}$, $\widehat{p}^{j}$ are the usual canonical
coordinates: $\left[  \widehat{q}^{i},\widehat{p}^{j}\right]  =i\hbar
\widehat{I}\delta^{ij}\ $. Here $\widehat{\mathbf{Q}}$ and $\widehat
{\mathbf{P}}$ are formally traceless operators: Tr$\left(  \widehat
{\mathbf{Q}}\right)  =$Tr$\left(  \widehat{\mathbf{P}}\right)  =0$. Next, we
determine $\widehat{\mathbf{Q}}\widehat{\mathbf{P}}$ and $\widehat{\mathbf{P}%
}\widehat{\mathbf{Q}}$:
\begin{equation}
\widehat{\mathbf{Q}}\widehat{\mathbf{P}}=\frac{1}{2}\times\frac{1}{3}\left(
\widehat{\mathbf{q}}\cdot\widehat{\mathbf{p}}-\widehat{\mathbf{p}}%
\cdot\widehat{\mathbf{q}}\right)  +\frac{i}{3\omega}\beta\times\left\{
\frac{\widehat{\mathbf{p}}^{2}}{2m_{0}}+\frac{m_{0}\omega^{2}}{2}%
\widehat{\mathbf{q}}^{2}+\frac{2\omega}{\hbar}\beta\mathbf{S}\cdot
\mathbf{L}\right\}  \equiv\frac{i\hbar}{2}\widehat{I}+\left(  \frac{i\beta
}{3\omega}\right)  \widehat{H}_{\text{SS}}\ ,
\end{equation}
where now
\begin{equation}
\widehat{H}_{\text{SS}}\equiv\widehat{H}_{\text{HO}}+\frac{2\omega}{\hbar
}\beta\mathbf{S}\cdot\mathbf{L}=\frac{\widehat{\mathbf{p}}^{2}}{2m_{0}}%
+\frac{m_{0}\omega^{2}}{2}\widehat{\mathbf{q}}^{2}+\frac{2\omega}{\hbar}%
\beta\mathbf{S}\cdot\mathbf{L\ }%
\end{equation}
is the Dirac Oscillator \cite{MH2,MOR,MH3,IT,CO,UI}, except for the additive
term $3m_{0}c^{2}$, in which $\widehat{H}_{\text{SS}}$ picks up a spin-orbit
term, with $\mathbf{L=}\widehat{\mathbf{q}}\times\widehat{\mathbf{p}}$ the
orbital angular momentum of $m_{0}$, and $S_{k}\mathbf{\equiv}\left(
1/2\right)  \Sigma_{k}=\left(  1/2\right)  \alpha_{i}\alpha_{j}\ $is the spin
of $m_{0}$, with $i,j,k$ cyclically. Correspondingly%
\begin{equation}
\widehat{\mathbf{P}}\widehat{\mathbf{Q}}=-\frac{i\hbar}{2}\widehat{I}-\left(
\frac{i\beta}{3\omega}\right)  \widehat{H}_{\text{SS}}\ .
\end{equation}
Therefore we now get%
\begin{equation}
\left[  \widehat{\mathbf{Q}},\widehat{\mathbf{P}}\right]  =i\hbar\left(
\widehat{I}+\left(  \frac{2}{3}\frac{\beta}{\hbar\omega}\right)  \widehat
{H}_{\text{SS}}\right)  ,\ \ \ \ \ \ \ \ \ \ \left\{  \widehat{\mathbf{Q}%
},\widehat{\mathbf{P}}\right\}  =\widehat{0}\ , \label{g20}%
\end{equation}
together with%
\[
\left[  \widehat{H}_{\text{HO}},\widehat{Q}^{j}\right]  =-i\frac{c^{2}}%
{\omega}\widehat{P}^{j}\ ,\ \ \ \ \ \ \ \ \ \ \left[  \widehat{H}_{\text{HO}%
},\widehat{P}^{j}\right]  =i\omega\left(  \frac{\hbar\omega}{c}\right)
^{2}\widehat{Q}^{j}\ ,
\]
in which $\widehat{\chi}=2\beta/3\hbar\omega$, and $\beta$ may be interpreted
as a `charge operator'. From its definition, $\widehat{H}\equiv-\left(
3/2\right)  \beta\widehat{H}_{\text{SS}}$ incorporates negative-energy
eigenvalues into the system. In fact, the structure of the first equation
(\ref{g20}) reminds us again of the Zitterbewegung\textit{\ }(ZB)\textit{\ }%
phenomenon On the other hand, note that, from the point of view of SUSYQM,
$\widehat{\mathbf{Q}}$ and $\widehat{\mathbf{P}}$ are SUSY charges \cite{WI}:%
\begin{equation}
\widehat{\mathbf{Q}}^{2}=\frac{c}{3\hbar\omega^{3}}\widehat{H}_{\text{SS}%
}\ ,\ \ \ \ \ \ \ \ \ \ \widehat{\mathbf{P}}^{2}=\frac{\hbar\omega}{3c^{2}%
}\widehat{H}_{\text{SS}}\ ,\nonumber
\end{equation}
together with the second equation (\ref{g20}). Here%
\begin{equation}
\widehat{H}_{\text{SS}}\equiv\frac{3}{2}\hbar\omega\left\{  \left(  \frac
{c}{\hbar\omega}\right)  ^{2}\widehat{\mathbf{P}}^{2}+\left(  \frac{\omega}%
{c}\right)  ^{2}\widehat{\mathbf{Q}}^{2}\right\}  =\frac{\widehat{\mathbf{p}%
}^{2}}{2m_{0}}+\frac{m_{0}\omega^{2}}{2}\widehat{\mathbf{q}}^{2}+\frac
{2\omega}{\hbar}\beta\mathbf{S}\cdot\mathbf{L}\ .
\end{equation}
Furthermore%
\begin{equation}
\left[  \widehat{H}_{\text{SS}},\widehat{\mathbf{Q}}\right]  =\left[
\widehat{\mathbf{Q}}^{2},\widehat{\mathbf{Q}}\right]  =\widehat{0}%
\ ,\ \ \ \ \ \ \ \ \ \ \left[  \widehat{H}_{\text{SS}},\widehat{\mathbf{P}%
}\right]  =\left[  \widehat{\mathbf{P}}^{2},\widehat{\mathbf{P}}\right]
=\widehat{0}\ , \label{g70}%
\end{equation}
which closes the algebra and confirms that $\widehat{H}_{\text{SS}}$ is
invariant under this supersymmetry. We note again that $H_{\text{SS}}\geq0$.
That is, the Hamiltonian has only non-negative eigenvalues. Let us suppose
that $\left\vert E_{a}\right\rangle $ is an eigenstate of $H_{\text{SS}}$ with
positive eigenvalue $E_{a}>0$. Then it follows that $\left\vert E_{a}%
\right\rangle ^{\prime}\varpropto\widehat{Q}\left\vert E_{a}\right\rangle $ is
also an eigenstate with the same positive eigenvalue. Relations (\ref{g70}),
together with $\left\{  \widehat{\mathbf{Q}},\widehat{\mathbf{P}}\right\}
=\widehat{0}$, is the graded algebra of a supersymmetric system consisting of
a relativistic spin-$\left(  1/2\right)  $ particle interacting with an
electric classical field \cite{BR2,BR3}. In the present case, the magnitude of
the electric field is proportional to $r\equiv\left\vert \widehat{\mathbf{q}%
}\right\vert =\left\vert \mathbf{q}\right\vert $ (the gradient of
$\mathbf{q}^{2}$).

\subsection{Dynamical group symmetries}

Some dynamical symmetries can be constructed for the harmonic oscillator in
$\left(  3+1\right)  $-dimensions. To this end, let us assume that the quantum
system is described by the vector-like $\widehat{q}^{i},\widehat{p}^{j}$
($i,j=1,2,3$) noncanonical `internal' coordinates. We assume that they satisfy
the commutation relations%
\begin{equation}
\left[  \widehat{q}^{i},\ \widehat{p}^{j}\right]  =i\hbar\delta^{ij}\left(
\widehat{I}+\frac{\chi}{\hbar\omega}\widehat{H}\right)  \equiv i\hbar
\delta^{ij}\widehat{\mathcal{O}}\ , \label{d10}%
\end{equation}
in which $\widehat{H}$ is the Hamiltonian of the system, $\chi$ a
dimensionless real constant, and $\omega$ the frequency of the oscillator. For
a harmonic oscillator we have to adjoin the commutation relations%
\begin{equation}
\frac{d\widehat{q}^{j}}{dt}=\frac{i}{\hbar}\left[  \widehat{H},\widehat{q}%
^{j}\right]  =\frac{\widehat{p}^{j}}{m_{0}}\ ,\ \ \ \ \ \ \ \ \frac
{d\widehat{p}^{j}}{dt}=\frac{i}{\hbar}\left[  \widehat{H},\widehat{p}%
^{j}\right]  =-m_{0}\omega^{2}\widehat{q}^{j}\ . \label{d12}%
\end{equation}

Now we look for a dynamical symmetry (a Lie algebra) associated with this
system. Jacobi's identity for the triad $\widehat{q}^{i},\widehat{p}%
^{i},\widehat{q}^{j}$ $\left(  i\neq j\right)  $ is
\begin{equation}
\left[  \left[  \widehat{q}^{i},\widehat{p}^{i}\right]  ,\widehat{q}%
^{j}\right]  +\left[  \left[  \widehat{q}^{j},\widehat{q}^{i}\right]
,\widehat{p}^{i}\right]  +\left[  \left[  \widehat{p}^{i},\widehat{q}%
^{j}\right]  ,\widehat{q}^{i}\right]  =\widehat{0},
\end{equation}
with no summation on the $i$ index. But from Eq.(\ref{d10}) $\left[  \left[
\widehat{p}^{i},\widehat{q}^{j}\right]  ,\widehat{q}^{i}\right]  =\widehat{0}%
$. Then
\begin{equation}
\left[  \left[  \widehat{q}^{i},\widehat{q}^{j}\right]  ,\widehat{p}%
^{i}\right]  =\left[  \left[  \widehat{q}^{i},\widehat{p}^{i}\right]
,\widehat{q}^{j}\right]  =i\hbar\left[  \widehat{\mathcal{O}},\widehat{q}%
^{j}\right]  \ .
\end{equation}
However, from Eq.(\ref{d12}), $\left[  \widehat{\mathcal{O}},\widehat{q}%
^{j}\right]  \neq\widehat{0}$:%
\begin{equation}
\left[  \widehat{\mathcal{O}},\widehat{q}^{j}\right]  =\frac{\chi}{\hbar
\omega}\left[  \widehat{H},\widehat{q}^{j}\right]  =-i\frac{1}{m_{0}}%
\frac{\chi}{\omega}\widehat{p}^{j}\neq\widehat{0}%
\end{equation}
only from the fact that $\chi\neq0$. This means that $\left[  \widehat{q}%
^{i},\widehat{q}^{j}\right]  \neq\widehat{0}$. Similarly, from the triad
$\widehat{p}^{i},\widehat{q}^{i},\widehat{p}^{j}$ we find that $\left[
\widehat{p}^{i},\widehat{p}^{j}\right]  \neq\widehat{0}$.

If we denote the generators of rotations for this system (i.e., the system's
angular momentum components) by $\widehat{J}^{k}$ , $k=1,2,3$, which satisfy
angular momentum commutation relations%
\begin{equation}
\left[  \widehat{J}^{i},\widehat{J}^{j}\right]  =i\hslash\varepsilon
^{ijk}\ \widehat{J}^{k}\ ,
\end{equation}
then
\begin{equation}
\left[  \widehat{J}^{i},\widehat{q}^{j}\right]  =i\hslash\varepsilon
^{ijk}\widehat{q}^{k}\ ,\ \ \ \ \ \ \ \ \left[  \widehat{J}^{i},\widehat
{p}^{j}\right]  =i\hslash\varepsilon^{ijk}\widehat{p}^{k}\ . \label{d15}%
\end{equation}
Next, the commutators $\left[  \widehat{q}^{i},\widehat{q}^{j}\right]  $ and
$\left[  \widehat{p}^{i},\widehat{p}^{j}\right]  $ transform as pseudo-tensors
under space rotations. This can be seen by using Jacobi's identity for the
triad $\left[  \left[  \widehat{q}^{i},\widehat{q}^{j}\right]  ,\widehat
{J}^{j}\right]  $ (no summation on $j$):%

\begin{equation}
\left[  \left[  \widehat{q}^{i},\widehat{q}^{j}\right]  ,\widehat{J}%
^{j}\right]  +\left[  \left[  \widehat{J}^{j},\widehat{q}^{i}\right]
,\widehat{q}^{j}\right]  +\left[  \left[  \widehat{q}^{j},\widehat{J}%
^{j}\right]  ,\widehat{q}^{i}\right]  =\widehat{0}\ . \label{d20}%
\end{equation}
Because $\left[  \widehat{q}^{j},J^{j}\right]  =\widehat{0}$, from
Eq.(\ref{d20}) we get%
\begin{equation}
\left[  \left[  \widehat{q}^{i},\widehat{q}^{j}\right]  ,J^{j}\right]
+\left[  \left[  J^{j},\widehat{q}^{i}\right]  ,\widehat{q}^{j}\right]
=\widehat{0},\ \ \ \ \ \ \ \ \left[  \left[  \widehat{p}^{i},\widehat{p}%
^{j}\right]  ,J^{j}\right]  +\left[  \left[  J^{j},\widehat{p}^{i}\right]
,\widehat{p}^{j}\right]  =\widehat{0}\ .
\end{equation}
Thus from Eq.(\ref{d15}) we obtain%
\begin{equation}
\left[  \left[  \widehat{q}^{i},\widehat{q}^{j}\right]  ,J^{j}\right]
=i\hslash\varepsilon^{ijk}\left[  \widehat{q}^{i},\widehat{q}^{j}\right]
,\ \ \ \ \ \ \ \ \left[  \left[  \widehat{p}^{i},\widehat{p}^{j}\right]
,J^{j}\right]  =i\hslash\varepsilon^{ijk}\left[  \widehat{p}^{i},\widehat
{p}^{j}\right]  \ . \label{d25}%
\end{equation}
The only pseudo-tensor available here is just $J^{k}$. Therefore the solutions
for the commutators $\left[  \widehat{q}_{i},\ \widehat{q}_{j}\right]  $ and
$\left[  \widehat{p}_{i},\ \widehat{p}_{j}\right]  $, which are compatible
with Eqs.(\ref{d15}-\ref{d25}), become
\begin{equation}
\left[  \widehat{q}^{i},\widehat{q}^{j}\right]  =i\zeta\epsilon^{ijk}%
J^{k}\ ,\ \ \ \ \ \ \ \ \left[  \widehat{p}^{i},\widehat{p}^{j}\right]
=i\xi\epsilon^{ijk}J^{k}\ , \label{d30}%
\end{equation}
where $\zeta$ and $\xi$ are real constants (with clearly identifiable dimensions).

On the other hand, for the triad $\widehat{\mathcal{O}},\widehat{q}%
^{i},\widehat{q}^{j}$ $\left(  i\neq j\right)  $ we have%
\begin{equation}
\left[  \left[  \widehat{q}^{i},\widehat{q}^{j}\right]  ,\widehat{\mathcal{O}%
}\right]  +\left[  \left[  \widehat{\mathcal{O}},\widehat{q}^{i}\right]
,\widehat{q}^{j}\right]  +\left[  \left[  \widehat{q}^{j},\widehat
{\mathcal{O}}\right]  ,\widehat{q}^{i}\right]  =\widehat{0}\ .
\end{equation}
As $\left[  \widehat{\mathcal{O}},\widehat{q}^{i}\right]  \varpropto
\widehat{p}^{i}$ , $\left[  \widehat{q}^{j},\widehat{\mathcal{O}}\right]
\varpropto\widehat{p}^{j}$ and $\left[  \widehat{q}^{i},\widehat{p}%
^{j}\right]  =\widehat{0}$ (for $i\neq j$), then $\left[  \left[  \widehat
{q}^{i},\widehat{q}^{j}\right]  ,\widehat{\mathcal{O}}\right]  =\widehat{0}$.
Hence, given the fact that $\left[  \widehat{q}^{i},\widehat{q}^{j}\right]
\varpropto\widehat{J}^{k}$, then
\begin{equation}
\left[  \widehat{\mathcal{O}},J^{k}\right]  =\left[  \widehat{H},J^{k}\right]
=\widehat{0}\ ,\ \ \ \ \ k=1,2,3.
\end{equation}
This means that the $J^{k}$ are conserved quantities.

At this stage of the problem, one wonders what sort of relations there are
among the various real parameters $\chi$,$\zeta$,$\xi$, etc., that appear in
the theory. To this end, let us choose the triad $\widehat{q}^{l}$%
,$\widehat{q}^{m}$,$\widehat{p}^{i}$. We then select Eqs.(\ref{d10}%
,\ref{d12},\ref{d15},\ref{d30}) to insert them (correspondingly) into Jacobi's
identity
\begin{equation}
\left[  \left[  \widehat{q}^{l},\widehat{q}^{m}\right]  ,\widehat{p}%
^{i}\right]  +\left[  \left[  \widehat{p}^{i},\widehat{q}^{l}\right]
,\widehat{q}^{m}\right]  +\left[  \left[  \widehat{q}^{m},\widehat{p}%
^{i}\right]  ,\widehat{q}^{l}\right]  =\widehat{0}\ ,
\end{equation}
we obtain%
\begin{equation}
\zeta\epsilon^{mlk}\epsilon^{kik^{\prime}}\widehat{p}^{k^{\prime}}-\hbar
\chi\delta^{il}\widehat{p}^{m}+\hbar\chi\delta^{mi}\widehat{p}^{l}=\widehat
{0}.
\end{equation}
If $m=1,$ $l=2$ , and then making $i=2$, we get $\zeta=-\chi/\omega$.

Following the same steps as above but for the triad $\widehat{p}^{l}%
,\widehat{p}^{m},\widehat{q}^{i}$, we find that $\xi=-\chi m_{0}\omega$.
Therefore, if $\chi\gtrless0$, then $\zeta\lessgtr0$ and $\xi\lessgtr0$. That
is to say, $\xi=m_{0}\omega^{2}\zeta$.

Hence the required commutation relations are%
\begin{equation}
\left[  \widehat{q}^{l},\widehat{q}^{m}\right]  =-i\frac{\chi}{\omega}%
\epsilon^{lmk}\widehat{J}^{k},\ \ \ \ \ \ \ \ \ \ \ \left[  \widehat{p}%
^{l},\widehat{p}^{m}\right]  =-i\frac{\chi\omega}{m_{0}}\epsilon^{lmk}%
\widehat{J}^{k}\ .
\end{equation}

To sum up, the (order 10, rank 2) Lie algebra generated by $\widehat{q}^{i}%
$,$\widehat{p}^{j}$,$\widehat{J}^{k}$, and $\widehat{H}$, is given by the
commutation relations%
\begin{equation}
\left[  \widehat{q}^{i},\widehat{p}^{j}\right]  =i\hbar\delta^{ij}\left(
\widehat{I}+\frac{\chi}{\hbar\omega}\widehat{H}\right)  \ ,\label{d40}%
\end{equation}%
\begin{equation}
\left[  \widehat{H},\widehat{q}^{j}\right]  =-i\hbar\widehat{p}^{j}%
\ ,\ \ \ \ \ \ \ \ \ \ \left[  \widehat{H},\widehat{p}^{j}\right]  =i\hbar
m_{0}\omega^{2}\widehat{q}^{j}\ ,
\end{equation}%
\begin{equation}
\left[  \widehat{q}^{i},\widehat{q}^{j}\right]  =-i\frac{\chi}{\omega}%
\epsilon^{ijk}J^{k}\ ,\ \ \ \ \ \ \ \ \ \ \ \left[  \widehat{p}^{i}%
,\widehat{p}^{j}\right]  =-i\frac{\chi\omega}{m_{0}}\epsilon^{ijk}\widehat
{J}^{k}\ ,
\end{equation}%
\begin{equation}
\left[  \widehat{J}^{i},\widehat{J}^{j}\right]  =i\hslash\varepsilon
^{ijk}\ \widehat{J}^{k}\ ,
\end{equation}%
\begin{equation}
\left[  \widehat{J}^{i},\widehat{q}^{j}\right]  =i\hslash\varepsilon
^{ijk}\widehat{q}^{k}\ ,\ \ \ \ \ \ \ \ \left[  \widehat{J}^{i},\widehat
{p}^{j}\right]  =i\hslash\varepsilon^{ijk}\widehat{p}^{k}\ ,
\end{equation}%
\begin{equation}
\left[  \widehat{H},\widehat{J}^{k}\right]  =\widehat{0}\ .
\end{equation}
Formally, if $\chi\neq0$, the central term $i\hbar\delta^{ij}\widehat{I}$ in
Eq.(\ref{d40}) can be reabsorbed into $\widehat{\mathcal{O}}$ and is
mathematically unimportant for a finite-dimensional semisimple Lie algebra.
However, we must distinguish the nature of the symmetry algebra according to
the values that the real parameter `$\chi$' can have. There are three possible
cases: (a) If $\chi\rightarrow0$ and $\omega\rightarrow0$ with $\left(
\chi/\omega\right)  \rightarrow\hbar/mc^{2}$, leading to the important case of
the Poincar\'{e} algebra in $(3+1)$-dimensions. Additionally, the case
$\chi\rightarrow0$ leads to the (isotropic) harmonic oscillator algebra in
3-dimensions, also known as the Newton-Hook algebra \cite{BRU}, the Heisenberg
algebra being a subalgebra of it. Here $i\hbar\delta^{ij}\widehat{I}$ is the
non-trivial center of the algebra and it cannot be eliminated. (b) If $\chi
<0$, this symmetry corresponds to the compact Lie algebra so(5). The
generators $\widehat{q}^{i}$,$\widehat{p}^{j}$,$\widehat{J}^{k}$, and
$\widehat{H}$ are traceless. This quantum model can be applied, for instance,
to the study of the electron Zitterbewegung\textit{\ }\cite{BA}\textit{. }In
this case then $\omega=\omega_{\text{Zitt}}=2m_{0}c^{2}/\hbar$. (c) If
$\chi>0$, the generated symmetry becomes the non-compact Lie algebra so(3,2).
This algebra can be studied, for example, from the point of view of Dirac's
representation \cite{DI}. This representation could be applied to the study of
some hadron resonances \cite{BR0} in the context of Regge trajectories. We
call to mind here some early works which describe different systems of
harmonic oscillators in terms of dynamical symmetry \cite{HAS,SC}. Actually,
there is an extensive (modern) literature on harmonic oscillator models based
on the dynamical algebra so(3,2) (see for instance Refs. \cite{AL1,GA,AL2,NA}).

These Lie algebras are well known, particularly so(5), so we are not going to
get involved at present into the discussion of the (energy) spectrum of
$\widehat{H}$ in terms of the eigenvalues of, say, $J_{3}$, etc. It is enough
to say that it is possible to find closed results, according to the
representation we use to describe the system in each case.

\subsection{Canonical coordinates}

Incidentally, we consider a general quantum system described by canonical
coordinates $\widehat{Q}^{\nu}$ and $\widehat{P}^{\mu}$ satisfying the
Heisenberg algebra \cite{BR1}
\begin{equation}
\left[  \widehat{P}^{\mu},\widehat{Q}^{\nu}\right]  =i\hbar\mathbb{I}g^{\mu
\nu}\ , \label{c10}%
\end{equation}
with the metric signature $g\left(  +---\right)  $, where $\mathbb{I\equiv
}I_{n\times n}\otimes I$ represents a $n$-block identity matrix such that we
may realize these operators in the general form$\ $%
\begin{equation}
\widehat{Q}^{\nu}=\widehat{\eta}\widehat{q}^{\nu}\equiv\widehat{\eta}%
\otimes\widehat{q}^{\nu}\ ,\ \ \ \ \ \ \ \ \ \ \widehat{P}^{\mu}=\widehat
{\eta}\widehat{p}^{\mu}\equiv\widehat{\eta}\otimes\widehat{p}^{\mu}\ ,
\label{c20}%
\end{equation}
where, in coordinate representation, $\widehat{p}^{\mu}=-i\hbar\partial
/\partial q_{\mu}$ . Here $\widehat{\eta}$ is a constant $n\times n$ Hermitian
matrix satisfying $\widehat{\eta}^{2}=I_{n\times n}.$ Thus we can define a
label $\left\vert \mathrm{Tr}\left(  \widehat{\eta}\right)  \right\vert $
associated with each representation of the Heisenberg algebra (\ref{c10}),
with $n\geq\left\vert \mathrm{Tr}\left(  \widehat{\eta}\right)  \right\vert
\geq0$. Representations satisfying $\left\vert \mathrm{Tr}\left(
\widehat{\eta}\right)  \right\vert =n$ correspond to the usual ones
$(\widehat{\eta}=I_{n\times n})$ where $Q^{\nu}$,$\ P^{\mu}$ are reducible
operators for $n\geq2$.

The Hilbert space is defined as $L^{2}(\mathbb{R}^{3})\otimes\mathbb{C}^{n}$.
It consists of $n$-component column vectors where each component $\psi_{i}$ is
a complex valued function of the $\left(  3+1\right)  $-dimensional (flat)
space-time coordinates $\mathbf{q,}t.$ The coordinates $\widehat{Q}^{i}$
$(i=1,2,3)$ consist of three self-adjoint operators. The momentum operator
$\widehat{P}^{j}=$ $-i\hbar\widehat{\eta}\partial/\partial q_{j}$ is defined
as the Fourier transformation of the position operator $\widehat{Q}^{j}$ .

Minimal interactions can now be introduced by means of the prescription
$\widehat{P}^{\mu}\rightarrow\widehat{P}^{\mu}-g\widehat{A}^{\mu}$, where $g$
is the coupling constant, $A_{\mu}$ is a gauge field $(\mu=0,1,2,3)$. Note
that here $\widehat{P}^{0}=$ $i\hbar I_{n\times n}\partial/\partial q_{0}$.
This is the basis of the so-called \textit{gauge principle} whereby the form
of the interaction is determined on the basis of local gauge invariance. The
covariant derivative $D^{\mu}\equiv(i/\hbar)\left(  \widehat{P}^{\mu
}-g\widehat{A}^{\mu}\right)  $ turns out to be of fundamental importance to
determine the field strength tensor of the theory. It will be the operator
which generalizes from electromagnetic-like interactions.

It is well known that the expression for the relativistic energy may be used
to form the classical free particle Hamiltonian. The analogous quantum
mechanical expression could be constructed by replacing the classical momentum
$\mathbf{p}$ with its quantum mechanical operator $\widehat{\mathbf{p}}$, with
components $\widehat{p}^{i}=-i\hbar\partial/\partial q_{i}$ , which produces
the spin-$0$ free particle wave equation
\begin{equation}
\left(  c^{2}\widehat{\mathbf{p}}^{2}+m_{0}^{2}c^{4}\right)  ^{1/2}\Phi\left(
q\right)  =c\widehat{p}^{0}\Phi\left(  q\right)  =i\hbar\frac{\partial
}{\partial t}\Phi\left(  q\right)  \ . \label{c30}%
\end{equation}
As is well known, this equation does not satisfy some of the conditions
required by special relativity. The wave equation is not covariant, and the
square root term introduces ambiguity. The Klein-Gordon equation (KGE) solves
both of these problems simply by taking the square of the original energy
expression and extending the result to a quantum mechanical wave equation:
\begin{equation}
\left(  g_{\mu\nu}\widehat{p}^{\mu}\widehat{p}^{\nu}-m_{0}^{2}c^{2}\right)
\Phi\left(  q\right)  =0\ , \label{c40}%
\end{equation}
where $\widehat{p}^{\mu}=-i\hbar\partial/\partial q_{\mu}$ \cite{GR}. The
resulting wave equation is covariant, but suffers from other problems.
Negative energy solutions to this equation are possible, which do not have a
readily obvious explanation. Besides, the probability density $\Phi^{\ast}%
\Phi$ fluctuates with time.

The expression in Eq. (\ref{c40}) is still valid if we make the replacement
$\widehat{p}^{\mu}=-i\hbar\partial/\partial q_{\mu}\rightarrow\widehat{P}%
^{\mu}=\sigma_{1}\widehat{p}^{\mu}\ $, with $\sigma_{j=1}$ a Pauli matrix, so
that%
\begin{equation}
\left(  m_{0}^{2}c^{4}-\widehat{P}^{j}\widehat{P}_{j}\right)  \Phi\left(
q\right)  =\left(  \widehat{P}^{0}\right)  ^{2}\Phi\left(  q\right)  =\left(
m_{0}^{2}c^{4}-\widehat{p}^{j}\widehat{p}_{j}\right)  \Phi\left(  q\right)
=\left(  \widehat{p}^{0}\right)  ^{2}\Phi\left(  q\right)  \ , \label{c50}%
\end{equation}
since $\left[  \widehat{p}^{i},\sigma_{j}\right]  =\widehat{0}$ and
$\sigma_{j}^{2}=I_{2\times2}$. Notice that $\Phi$ is now a two-component wave
function, that is to say, the Hilbert space has been enlarged. Pauli matrices
$\sigma_{j}$, satisfy the well-known relations%
\begin{equation}
\left\{  \sigma_{i},\ \sigma_{j}\right\}  =2\delta_{ij},\qquad\sigma_{j}%
^{2}=I_{2\times2},\qquad i,j,k=1,2,3,\ \ \sigma_{i}\sigma_{j}=i\sigma
_{k},\ \text{cyclically .} \label{c60}%
\end{equation}

Next we briefly discuss the interaction of a basic effective minimal coupling
between spinless quarks \cite{PE}. Let%
\begin{align}
\widehat{P}^{0}  &  \rightarrow\widehat{\Pi}^{0}=i\hbar c\sigma_{1}\nabla
^{0}-\left(  \sigma_{2}\sqrt{\frac{-a_{1}^{2}}{r}}+a_{2}\sigma_{3}\sqrt
{r}\right)  \ ,\label{c70}\\
\widehat{P}^{k}  &  \rightarrow\widehat{\Pi}^{k}=i\hbar\sigma_{1}\nabla
^{k}-a_{3}\sigma_{2}\frac{q^{k}}{r}\ , \label{c71}%
\end{align}
be a minimal replacement in the KGE, with $a_{1}$, $a_{2}$,$\ a_{3}$, real
constants. If we introduce the minimal coupling $\widehat{P}^{\mu}%
\rightarrow\widehat{\Pi}^{\mu}$ into Eq.(\ref{c50}) and divide the resulting
equation by $2m_{0}c^{2}$, we get%
\begin{equation}
\left\{  -\frac{\hbar^{2}}{2m_{0}}\mathbf{\nabla}^{2}-\left(  \frac{a_{1}^{2}%
}{2m_{0}c^{2}}-2\hbar a_{3}\sigma_{3}\right)  \frac{1}{r}+\frac{a_{2}^{2}%
}{2m_{0}c^{2}}r\right.  \label{c80}%
\end{equation}%
\[
\left.  +\frac{1}{2}m_{0}c^{2}\left(  1+\left(  \frac{a_{3}}{m_{0}c^{2}%
}\right)  ^{2}\right)  \right\}  \Phi\left(  q\right)  =-\frac{\hbar^{2}%
}{2m_{0}}\frac{\partial^{2}}{\partial t^{2}}\Phi\left(  q\right)  .
\]
Note that $\widehat{\Pi}^{0}$ is not Hermitian. However when placed into the
KGE, each piece become in fact Hermitian. This feature also takes place in the
Dirac Oscillator\textit{\ }definition\textit{\ }\cite{MH2}. Rearranging terms
in Eq.(\ref{c80}) we obtain the relativistic wave equation
\begin{equation}
\left(  -\frac{\hbar^{2}}{2m_{0}}\mathbf{\nabla}^{2}-\frac{\alpha}%
{r}+kr\right)  \Phi\left(  \mathbf{q}\right)  =\epsilon_{KG}\Phi\left(
\mathbf{q}\right)  \ , \label{c90}%
\end{equation}
with
\begin{equation}
\epsilon_{KG}\equiv\frac{1}{2}m_{0}c^{2}\left(  \left(  \frac{E_{KG}}%
{m_{0}c^{2}}\right)  ^{2}-\left(  \frac{a_{3}}{m_{0}c^{2}}\right)
^{2}-1\right)  \ ,\qquad\alpha\equiv\frac{a_{1}^{2}}{2m_{0}c^{2}}-2\hbar
a_{3}\sigma_{3}\ ,\qquad k\equiv\frac{a_{2}^{2}}{2m_{0}c^{2}}\ . \label{c100}%
\end{equation}

Of course, in this example we are not dealing with electromagnetic-like
interactions. Notice that Eq.(\ref{c80}) has been obtained (as a minimal
coupling) within the framework of a relativistic equation. Furthermore, for
the free particle ($a_{1}=a_{2}=a_{3}=0$) the limit $\left\vert \widehat
{\mathbf{p}}\right\vert \rightarrow0$ yields%
\begin{equation}
\epsilon_{KG}=\frac{1}{2}m_{0}c^{2}\left(  \left(  \frac{E_{KG}}{m_{0}c^{2}%
}\right)  ^{2}-1\right)  \rightarrow\frac{1}{2}m_{0}c^{2}\left(  \left(
\frac{\left(  c^{2}\widehat{\mathbf{p}}^{2}+m_{0}^{2}c^{4}\right)  ^{1/2}%
}{m_{0}c^{2}}\right)  ^{2}-1\right)  \rightarrow\frac{\widehat{\mathbf{p}}%
^{2}}{2m_{0}}\ , \label{c110}%
\end{equation}
as one should anticipate. Thus we reacquire the nonrelativistic expression for
the kinetic energy.

We also have that%
\begin{align}
F^{0k}  &  \varpropto\left[  \widehat{\Pi}^{0},\ \widehat{\Pi}^{k}\right]
=-i\hbar\left(  a_{2}\sigma_{2}+\frac{1}{2}\frac{a_{1}}{r^{3/2}}\sigma
_{3}\right)  \widehat{q}^{k}\ ,\label{c120}\\
F^{ij}  &  \varpropto\left[  \widehat{\Pi}^{i},\ \widehat{\Pi}^{j}\right]
=-ia_{3}\frac{\sigma_{3}}{r}L_{k}\ ,\quad i,j,k\text{ cyclically\ .}
\label{c121}%
\end{align}
Notice as well that in this case
\begin{equation}
\left[  A^{0},\ A^{k}\right]  =-2ia_{2}a_{3}\frac{q^{k}}{\sqrt{r}}\sigma
_{1}\neq\widehat{0}\ , \label{c130}%
\end{equation}
so that the vector potential $A^{\mu}$ is not Abelian. In QCD the strong color
field is mediated by massless vector bosons. Hence, the potential might be
expected to be of the Coulomb form. At large $r$, the quarks are subjected to
confining forces. It is found that the potential at large $r$ is linear
\cite{PE}. We are not going to solve here the eigenvalue problem (\ref{c90}),
since it has been widely studied in the context of the (non-relativistic)
Schr\"{o}dinger equation \cite{LA}. Notice that this basic model by no means
solves the general problem of, for instance, heavy quarkonia in the
relativistic quark model \cite{EB}.

\section{Conclusions}

This article deals with some possible applications of particular harmonic
oscillators-like models to particle physics in the framework of certain
noncanonical commutation relations. We have presented specific types of
harmonic oscillators in order to study mathematical relations among
well-established symmetries in physics such as supersymmetry, quark
interactions, and particle-antiparticle symmetries. It remains to be given a
more solid foundation for this portrayal. For instance, if one wants to be
meaningful, there is a need (a difficult one) to incorporate various degrees
of freedom existing in meson-quark physics, accounting for physical effects
such as retardation and radiative corrections, amongst others \cite{EB}.
Perhaps one way of particularly doing this is by formally generalizing
Eq.(\ref{i1}) as%
\begin{equation}
\left[  \widehat{q}^{i},\widehat{p}^{j}\right]  =i\hbar\delta^{ij}\left(
\widehat{I}+f\left(  \widehat{O}_{1},...\widehat{O}_{n}\right)  \right)  \ ,
\label{c140}%
\end{equation}
where $f\left(  \widehat{O}_{1},...\widehat{O}_{n}\right)  $ is a power series
function of the constant $n$-compatible obserbables $\widehat{O}_{j}$ of the
system, i.e., they commute with the Hamiltonian and with each other. Actually,
at present our research is directed to extract information from Eq.(\ref{c140}%
) for simple systems in $\left(  3+1\right)  $-dimensions, where a somehow
judicious form for the function $f$ is to be set. This work was supported by
Direcci\'{o}n de Investigaci\'{o}n, Universidad de Concepci\'{o}n, Chile,
through grant P.I. 207.011.046-1.0.

\end{document}